\documentstyle[psfig]{article}
\begin{document}
 \pagestyle{plain}
 \newcommand{\adspb}{\vspace{-20pt}}
 \newcommand{\qqq}[1]{{\textstyle #1}}
 \newlength{\przeditem}
 \newlength{\poitem}
 \setlength{\przeditem}{-15pt}
 \setlength{\poitem}{-5pt}
 \newlength{\mimi}
 \setlength{\mimi}{7cm}
 \newcommand{\odstext}{\renewcommand{\baselinestretch}{2.0}}
 \newcommand{\odsrys}{\renewcommand{\baselinestretch}{1.}}
 \odstext \

 \begin{center}
 {\Large \bf
 Measurable Difference in Cherenkov Light \\ 
 between Gamma and Hadron Induced EAS.}\\

 \vspace{1.0cm}
 Herv\'{e} Cabot$^{1}$,
 Christian Meynadier$^{1}$,
 Dorota Sobczy\'{n}ska$^{2}$,\\
 Barbara Szabelska$^{3}$, Jacek Szabelski$^{1,3}$,
 Tadeusz Wibig$^{2}$\\

 \vspace{1cm}
 \parbox{11cm}{
 $^{1}$Universit\'{e} de Perpignan, Groupe de Physique Fondamentale,\\
 $\rule{1cm}{0cm}$ 52 av. Villeneuve, 68860 Perpignan Cedex, France\\
 $^{2}$University of {\L }\'{o}d\'{z}, Experimental Physics Dept.,\\
 $\rule{1cm}{0cm}$ ul.~Pomorska 149/153, 90-236 {\L }\'{o}d\'{z}, Poland\\
 $^{3}$Soltan Institute of Nuclear Studies, 90-950 {\L }\'{o}d\'{z}, 
 Box 447, Poland\\
 }

 \vspace{1cm}
 {\large \bf Abstract}
 
 \end{center}

 \hspace{1cm}
 \parbox{13cm}{
 We describe the possibly measurable difference
 in the Cherenkov light component\\ 
 of EAS induced by an electromagnetic particle (i.e. e$^{+}$, e$^{-}$ 
 or $\gamma$) and induced by\\ 
 a hadron (i.e. proton or heavier nuclei) in TeV range.\\
 The method can be applied in experiments
 which use wavefront sampling method\\ 
 of EAS Cherenkov light detection (e.g. THEMISTOCLE, ASGAT).\\
 }
 
 \newpage
 \section{
 Introduction
 }
 
 \noindent
 We refer to the Extensive Air
 Shower (EAS) experiments which are detecting cosmic rays (CR)
 in TeV energy range and are using Cherenkov light
 signal.   
 The primary goal is to identify
 and measure gamma ray flux from astrophysical
 objects. 
 Positive results obtained by 
 the Whipple Collaboration 
       \cite[{\em Vacanti et al., 1988}]{Whipple}
 and by 
    the THEMISTOCLE Collaboration 
       \cite[{\em Baillon et al., 1993}]{THEMISTOCLE}
 are encouraging. 
 (For the review of current status see
  vol. 1 of {\em the Proceedings of XXIV ICRC, Roma}, 1995,
  or {\em the Proceedings of the Padova Workshop on 
  TeV Gamma~--~Ray Astrophysics \lq \lq Towards a Major
  Atmospheric Cherenkov Detector--IV", Sept.~11--13, 1995,
  ed.~M.~Cresti}).\\
 There are two different methods currently used in this
 area: \lq \lq imaging" and \lq \lq wavefront sampling".
 We refer here to the \lq \lq wavefront sampling" method, which detects
 the Cerenkov light simultaneously in number 
 of places distributed on the field of the size 
 larger than 200 m.
 The timing and the signal
 amplitudes are used to identify and classify the event.
 (The method has been developed and is used by
  the THEMISTOCLE and ASGAT groups and is planned to be used
  by the CELESTE group at the Themis site in 
  the French Pyrenees).\\
 The main problem is to distinguish the gamma ray induced
 events from the larger background of proton induced showers,
 which is very difficult to achieve by examining the EAS
 on the Cherenkov light cone shape and the signal
 amplitudes in each detector. The DC signal
 from the Crab nebula direction has been found (only)
 due to very good angular resolution ($\approx$~2.5~mrad),
 tracking the source on the sky 
       \cite[{\em Baillon et al., 1993}]{THEMISTOCLE}.
 Thus the increase in number of events was  
 obtained when the detectors where centred on the source
 direction as compared with the off--source measurements
 (no identification of nature of primary particle
 was applied).\\

 \noindent
 Here we pay attention to the Cherenkov light produced
 by muons. We demonstrate that the Cherenkov light from muon
 observed at the distance of few tens of meters
 from the EAS core can arrive several nanoseconds before
 the \lq \lq main" signal produced by electrons and positrons.
 Identification of \lq \lq muon" signal can be used as
 identification of hadronic origin of parent energetic CR particle.
 At TeV energy range the ratio of number of muons in proton showers
 to the number of muons in electromagnetic (E--M) showers
 is about 100, if calculated for similar Cherenkov light
 intensity at 50~m from EAS core 
 (i.e. few muons with $E_{\mu}~>$~10~GeV in E--M showers,
 and few hundred muons in proton showers).
 In this paper we present results of theoretical analysis
 of the problem (approximate calculations and results
 of the Monte Carlo simulation) and we address the hardware 
 requirements which should be met for experimental
 identification of that effect.
 
 \vspace{-10pt}
 \section{
 Approximate calculations
 }
 To demonstrate how Cherenkov light produced by 
 the muon can come before the light produced by
 electrons and positrons it could be useful
 to perform approximate calculations and 
 clarify the picture of the event.\\
 The Cherenkov light is emitted when a charged particle
 is passing through the air (or other matter) 
 with velocity greater than local phase velocity of light. 
 The velocity of light is equal to
 $v_{air} \, = \, c / n$, where $c$ is the speed of light
 in vacuum and $n$ is the local refraction index.\\
 We are interested in the observations at the lateral 
 distance of more than 50~m from the EAS core 
 (crossing point of CR particle trajectory and ground surface).
 Most of E--M particles of EAS at TeV energies
 are closer to the CR particle trajectory and
 the E--M cascade develops few kilometers above
 the ground, not reaching the level 2~km~a.s.l.,
 and most of Cherenkov light is emitted by e$^{+}$/e$^{-}$
 at altitudes 3~--~12~km~a.s.l.\\
 Muons with energy above 5~GeV are faster than light in the
 atmosphere at sea level. They can produce Cherenkov
 radiation near to the detector. This light can be detected
 earlier than the Cherenkov light from E--M cascade.\\
 Muons are decay products of hadrons originated near to
 CR particle trajectory. Energetic muons can reach the
 observation level and can generate Cherenkov radiation
 near to the detector. The energy threshold for muon 
 to produce Cherenkov radiation in the atmosphere is
 about 5~GeV, so such muons are relativistic, and
 they might not decay flying several kilometers.\\
 We compare the time of flight of light emitted at the level $h$
 reaching the surface at distance $r$ from the EAS core
 with the time of flight of muon with energy $E_{\mu}$ which was
 emitted at the same altitude $h$ and reaching the same place
 at the distance $r$.
 We set the time reference point $t_{0}$ at the time of $h/c$.

 \noindent
 \noindent
 \noindent
 \parbox{9.5cm}{
 $l$ is the distance travelled by light and muon 
 ($l=\sqrt{r^{2}+h^{2}}$).\\
 
 For muons:\\
 \[ 
 \begin{array}{rcl}
 \Delta t_{\mu} & =  & - \frac{\textstyle h}{\textstyle c} 
  + \frac{\textstyle l}{\textstyle v_{\mu}} \\
 & & \\
 v_{\mu} & = & c \cdot \beta_{\mu} \\
 & & \\
 \beta_{\mu} & = & \sqrt{1 - 
 \frac{\textstyle 1}{\textstyle \gamma_{\mu}^{2}}} \\
 & & \\
 \gamma_{\mu} & = & \frac{\textstyle E_{\mu}}{\textstyle m_{\mu} c^{2}} \\
 & & \\
 E_{\mu} & >> & m_{\mu} c^{2}\\
 & & \\
 \frac{1}{\qqq{\beta_{\mu}}} & \approx & 1 + 
 \frac{\textstyle 1}{\textstyle 2} 
 \left( \frac{\textstyle m_{\mu} c^{2}}{\textstyle E_{\mu}}\right)^{2} +
 \frac{\textstyle 3}{\textstyle 8} 
 \left( \frac{\textstyle m_{\mu} c^{2}}{\textstyle E_{\mu}}\right)^{4} \\
 & & \\
 \Delta t_{\mu} & = & - \frac{\textstyle h}{\textstyle c} +
 \frac{\textstyle \sqrt{\textstyle r^{2}+h^{2}}}{\textstyle c} \cdot
 \left[ 1 +
 \frac{\textstyle 1}{\textstyle 2} 
 \left( \frac{\textstyle m_{\mu} c^{2}}{\textstyle E_{\mu}}\right)^{2} +
 \frac{\textstyle 3}{\textstyle 8} 
 \left( \frac{\textstyle m_{\mu} c^{2}}{\textstyle E_{\mu}}\right)^{4} 
 \right]
 \end{array}
 \]
 }
 \hfill
 \begin{minipage}{6cm}
 \setlength{\unitlength}{1mm}
 \begin{picture}(60,100)
 \put(10,10){\line(0,1){80}}
 \put(10,10){\line(1,0){45}}
 \put(10,76){\vector(1,-2){33}}
 \put(4,70){\makebox(0,0)[bl]{\it h}}
 \put(40,4){\makebox(0,0)[bl]{\it r}}
 \put(28,45){\makebox(0,0)[bl]{\it l}}
 \end{picture}

 \end{minipage}
 \\

 \adspb\ 
 
 \noindent
 For light it is necessary to take into calculation
 the variation of refraction index with the altitude.
 For the form: $n = 1 + \eta$ the approximate dependence
 is valid for these calculations
 \cite[{\em Hillas, 1982}]{Hillas}:
 $\eta = \eta_{0} \, \cdot \, exp(-h/h_{0})$, where
 $\eta_{0}$=2.9$\cdot$10$^{-4}$, $h_{0}$=7.1~km.
 \[ 
 \begin{array}{rcl}
 \Delta t_{l} & =  & - \frac{\textstyle h}{\textstyle c} + 
 \frac{\textstyle 1}{\textstyle c} 
 {\displaystyle
 \int_{\qqq{0}}^{\qqq{\sqrt{r^{2}+h^{2}}}} 
 \, (1+\eta) \, dx
 } \\
 & & \\
 \eta & = & \eta_{0} \, \cdot \, exp\left( 
 \frac{\textstyle -x}{\textstyle h_{0}} 
 \frac{\textstyle h}{\textstyle \sqrt{\textstyle r^{2}+h^{2}}}\right) \\
 & & \\
 \Delta t_{l} & =  & \frac{\textstyle 1}{\textstyle c} 
 \left\{\textstyle  \sqrt{\textstyle r^{2}+h^{2}} - h +
 \frac{\textstyle h_{0}}{\textstyle h} 
 \, \sqrt{\textstyle r^{2}+h^{2}} \cdot \eta_{0} \cdot
 \left[ 1 - exp\left( \frac{\textstyle -h}{\textstyle h_{0}} 
 \right) \right] \right\} 
 \end{array}
 \]
 \\

\begin{figure}[tb]
 \psfig{file=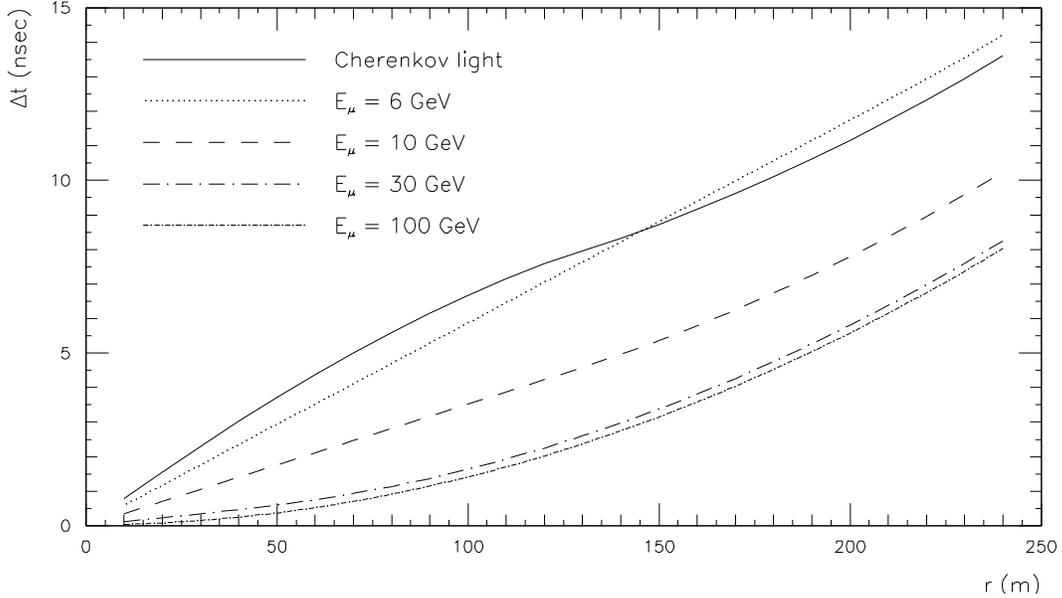,width=14.5cm,height=8.0cm}
 \odsrys \ 
 \caption
 {\label{fig:delta_t}
 The approximate relative time differences 
 $\Delta t_{l}$ and $\Delta t_{\mu}$ for Cherenkov
 light and muon.
 $t_{0}$ is the arrival time of
 ultra\-relativitic particle at the EAS core.
 $\Delta t$ is the time with respect to $t_{0}$.
 It is shown that energetic muons, and Cherenkov light produced 
 by them, can precede Cherenkov light produced by 
 $e^{+}$, $e^{-}$ of EAS by several nanoseconds.
 }
 \end{figure}

 \adspb\ 
 \noindent
 Using above formulae for light and muons we evaluate 
 for a given distance $r$ the minimum $\Delta t$
 varying the height of production $h$,
 however limited to 12~km. That limit has a physical 
 justification, since most of EAS of these energies
 are well developed
 above 2~km  (820~g/cm$^{2}$) and below 12~km~a.s.l. (195~g/cm$^{2}$).
 Without the limit the $\Delta t_{\mu}$
 is still smaller for higher energy muons.
 Results are compiled 
 in the Figure~\ref{fig:delta_t}.

 \odstext \

 \noindent
 We would like to underline two features presented
 in the Figure~\ref{fig:delta_t}:\\
 \vspace{\przeditem}
 \begin{itemize}
 \setlength{\itemsep}{-2pt}
 \item $\Delta t_{l}$ (represented by solid line) is nearly linearly
   proportional to the distance $r$;
 \item muons with energy above 10~GeV at the distances $r > 100~m$
   can arrive 3~nsec before the Cherenkov light of e$^{+}$/e$^{-}$ 
   origin; for $E_{\mu} > 30~GeV$ it could be up to 6~nsec
   difference.
 \end{itemize}
 \vspace{\poitem}
 We are going to discuss the second point in some details
 in following sections
 since this feature can help to distinguish between gamma
 and hadron induced events.\\
 Here we have listed some simplifications used in the
 approximate calculations presented in this work
 (Figure~\ref{fig:delta_t}). In our opinion
 they do not have large influence on $\Delta t$,
 however it is worth to know better the picture of
 real events.\\
 \vspace{\przeditem}
 \begin{itemize}
 \setlength{\itemsep}{-2pt}
 \item[--] All charged particles with velocity $v > c/n$
   ($n$ -- refraction index), can produce Cherenkov radiation.
   In the atmosphere (since $n$ is altitude dependent)
   the minimal energy for e$^{+}$/e$^{-}$ is 21.2~MeV at sea level
   and about 40~MeV at 10~km, and for muons: 5~GeV at sea level,
   and 10~GeV at 10~km.
 \item[--] Mean lifetime for muons is 
   $\tau$~=~2.2~$\cdot$~10$^{-6}$~sec
   which corresponds to 31~km for 5~GeV muon.
   Some muons can decay and the probability of decay 
   depends on energy.
 \item[--] Cherenkov light is not emitted in the direction 
   of charged particle, but at the angle $\theta_{c}$
   to the particle trajectory (surface of the cone)
   (cos($\theta_{c}$) = 1/($\beta \cdot n$), where $\beta$~=~$v$/c).
   For relativistic particles ($\gamma~>$~100) in the atmosphere 
   $\theta_{c}~\approx$~20~mrad at the sea level and
   $\theta_{c}~\approx$~12~mrad at about 12~km.
   These values are comparable with the acceptance
   angle of many Cherenkov light EAS detectors.
   Therefore the combined trajectory: charged particle
   and emitted light do not make a straight line and
   $\Delta t$ evaluation
   and acceptance geometry are complicated.
 \item[--] There are large fluctuations in EAS development.
   Very important role plays the level of the first
   interaction of CR particle. Normally, if first interaction
   is deeper in the atmosphere more Cherenkov light is produced,
   because the E--M cascade develops in volume of larger
   refraction index $n$. For muons the situation is opposite:
   we expect more muons if the first interaction takes place
   higher in the atmosphere, because in low density media
   the decay of meson to muon is more likely.
 \item[--] E--M cascade has lateral spread of tens of meters.
   Cherenkov photons can travel different distance
   than assumed in our approximate calculations and
   the $\Delta t_{l}$ might be a little smaller due
   to this geometry.
 \item[--] Some of Cherenkov photons can be scattered or absorbed
   in the atmosphere, and therefore are lost. 
   Atmospheric transmission due to Rayleigh scattering,
   aerosol (Mie) scattering and absorption in ozone
   are discussed in our earlier works
   \cite[{\em Attallah et al., 1995}]{Czer_I}.
 \item[--] Let $E_{CR}$ be the energy of CR particle which
   collides with nucleon of air nucleus. 
   The center of mass reference frame
   has $\gamma_{cm}~\approx~\sqrt{E_{CR}/2~GeV}$.
   If particle produced in collision would be pion 
   ($m_{\pi} = 0.14~GeV$)
   with Feynman's $x_{F}$~=~0
   then it would have in laboratory reference frame
   energy $E_{\pi}$ = $\sqrt{E_{CR}/2~GeV} \cdot m_{\pi}$,
   i.e. $E_{\pi}$ = 10~GeV for $E_{CR}$ = 10~TeV, or
   $E_{\pi}$ = 3~GeV for $E_{CR}$ = 1~TeV.
   So energetic muons are decay products of mesons
   produced in forward cone ($x_{F}~>$~0) in first and
   subsequent interactions.
 \item[--] When the primary CR particle is an E--M one
   (i.e. e$^{+}$, e$^{-}$ or $\gamma$), muons can be
   produced via photoproduction of hadrons.
   The cross section for such a process is
   $\sigma_{\gamma \rightarrow hadrons}~\approx$~100~$\mu$barn. 
   However the hadron (and then muon) energies
   are naturally smaller, than in corresponding
   case when the primary CR particle is a hadron.
 \end{itemize}
 
 \section{
 Monte--Carlo simulations.
 }
 The detailed analysis of the arrival time
 of Cherenkov light from E--M component and
 from muons was performed by Monte--Carlo
 simulations of EAS development in the atmosphere,
 light transmission, mirror reflection,
 photomultiplier efficiency and signal convolution.\\
 To simulate EAS development in the atmosphere
 we used the CORSIKA code v.~4.50 
 \cite[{\em Capdevielle et al., 1992}]{CORSIKA_I}
 \cite[{\em Knapp and Heck, 1995}]{CORSIKA_II}.
 This is a current version of detailed simulation program for EAS
 developed in Forschungszentrum Karlsruhe (Germany).
 It uses FORTRAN77, and different 
 CMZ \cite[{\em CodeME. S.A.R.L., 1993}]{CMZ}
 selections
 give the versions for different computer systems
 (IBM~M3090, VMS, DEC-Unix, APPLE Macintosh and
 transputers used in FZK).
 The DEC-Unix option has been adapted to Alpha~XP
 computers using Digital FORTRAN at Perpignan,
 and GNU's f2c converter at Alpha~XP, PC~486 with Linux,
 and PC~486 with DOS in {\L }\'{o}d\'{z}.
 The CORSIKA program has been used to study different aspects
 of Cherenkov radiation in EAS. Results presented in this
 work are extracted for further processing from
 simulations of 10 showers generated by vertical 6~TeV CR gamma
 and 10 showers by 10~TeV CR proton. We used the program
 option with GHEISHA code for low energy interactions,
 EGS electromagnetic interaction code, the Cherenkov  
 light \lq bunch size' of 1~photon per bunch,
 and CORSIKA Coulomb scattering parameter STEPFC~=~0.2.
 The primary CR particle energies had been selected
 to give similar Cherenkov photon (300--450~nm, no losses)
 densities at 50~m from EAS core, equal to about 1000~ph/m$^{2}$.
 The simulation time using Alpha~XP 175~MHz station
 was about 60~min. per gamma shower and about 40~min.
 per proton shower. The total (10 EAS) CORSIKA output
 have more than 105~Mb (for gammas) 
 and more than 80~Mb (for protons) 
 (mostly due to Cherenkov photon information,
 despite the fact that only photons pointed to
 preselected detector areas, 729~m$^{2}$ in total, 
 were memorized; 28 bytes per \lq bunch').\\

 \noindent
 The standard CORSIKA output for Cherenkov light 
 contains seven 4--byte real numbers per registered
 Cherenkov light bunch. These provide information on 
 number of Cherenkov photons in a bunch, 
 x~position of bunch at registration level, 
 y~position, 
 direction cosine to the x--axis, 
 direction cosine to the y--axis,
 altitude of bunch production,
 time of bunch arrival with respect to the time of the
      first interaction in the EAS.
 The program produces the bunch of Cherenkov photons,
 all in the same direction (on the cone surface), 
 instead of generating each photon at the cone surface.
 The average bunch size can be selected as an input parameter.
 In the default the photons have wavelength range
 300~--~450~nm.\\
 For further analysis we have prepared a system
 of relatively smaller programs for processing the Cherenkov
 light CORSIKA output. These programs rescale the
 photon wavelength range, apply the losses of photons in the 
 atmosphere (which are wavelength, altitude and pathlength dependent;
 see \cite[{\em Attallah et al., 1995}]{Czer_I} for more details.
 These programs calculate the detector mirror reflection
 probability according to
 \cite[{\em Riera and Espigat, 1994}]{Riera}
 (for the THEMISTOCLE mirrors).
 The photomultiplier quantum efficiency 
 for bialkali cathode was taken from
 \cite[{\em Philips Handbook, 1990}]{PhilipsHand}
 as for phototube PHILIPS~XP2020.\\
 The progams use HBOOK procedures of CERN's {\em packlib}
 \cite{HBOOK}
 and program PAW
 \cite{PAW}
 was used to make most of presented graphics.
 
 \vspace{-10pt}
 
 \vspace{-10pt}
 \section{
 Example.
 }
 To demonstrate the role of muons in Cherenkov light
 detection as result of Monte--Carlo simulations
 we present an example of one shower observed
 by one detector in one place. 
 EASs almost never look 'the same'
 because of fluctuations in their development
 and stochastic nature of most of physical processes
 involved. 
 Average, mean nor \lq most probable' 
 values do not describe well the situation
 we are going to present.
 Many features 
 are detector dependent and full detailed analysis
 should be performed for specific experimental setup.\\

 \noindent
 In our example we present result of Monte--Carlo
 simulation of EAS triggered by vertical CR proton
 with energy 10~TeV. The detector is at the 1650~m~a.s.l.,
 altitude of Themis site (French Pyrenees) 
 or Baksan Neutrino Observatory (Russia), where the Karpet detector
 is now equipped with Cherenkov device.
 Our simulated detector is located in ($x_{\qqq{d}}$,$y_{\qqq{d}}$) 
 position equal to (--50~m, 0~m). We will consider two detector areas:
 3~m~x~3~m and the circle of the area of 0.44~m$^{2}$
 both centred at ($x_{\qqq{d}}$,$y_{\qqq{d}}$).
 0.44~m$^{2}$ is the effective area of each of the THEMISTOCLE
 experiment detectors
 \cite[{\em Baillon et al., 1993}]{THEMISTOCLE}, 
 and heliostat mirrors in the CELESTE experiment would
 have effective area of $\approx$ 31~m$^{2}$ 
 (of 54~m$^{2}$ for each mirror)
 \cite[{\em Dumora et al., 1996}]{Celeste}).\\

 \noindent
 The muon of energy 88.52~GeV has been found at
 position (--53.07~m, --3.59~m), 
 with direction cosines
 (cos($\alpha_{\qqq{x}}$),cos($\alpha_{\qqq{y}}$)) equal to
 (--0.00196,0.00088) at $\Delta t_{\mu}$~=~0.3~nsec.\\
 \begin{figure}[tb]
 \psfig{file=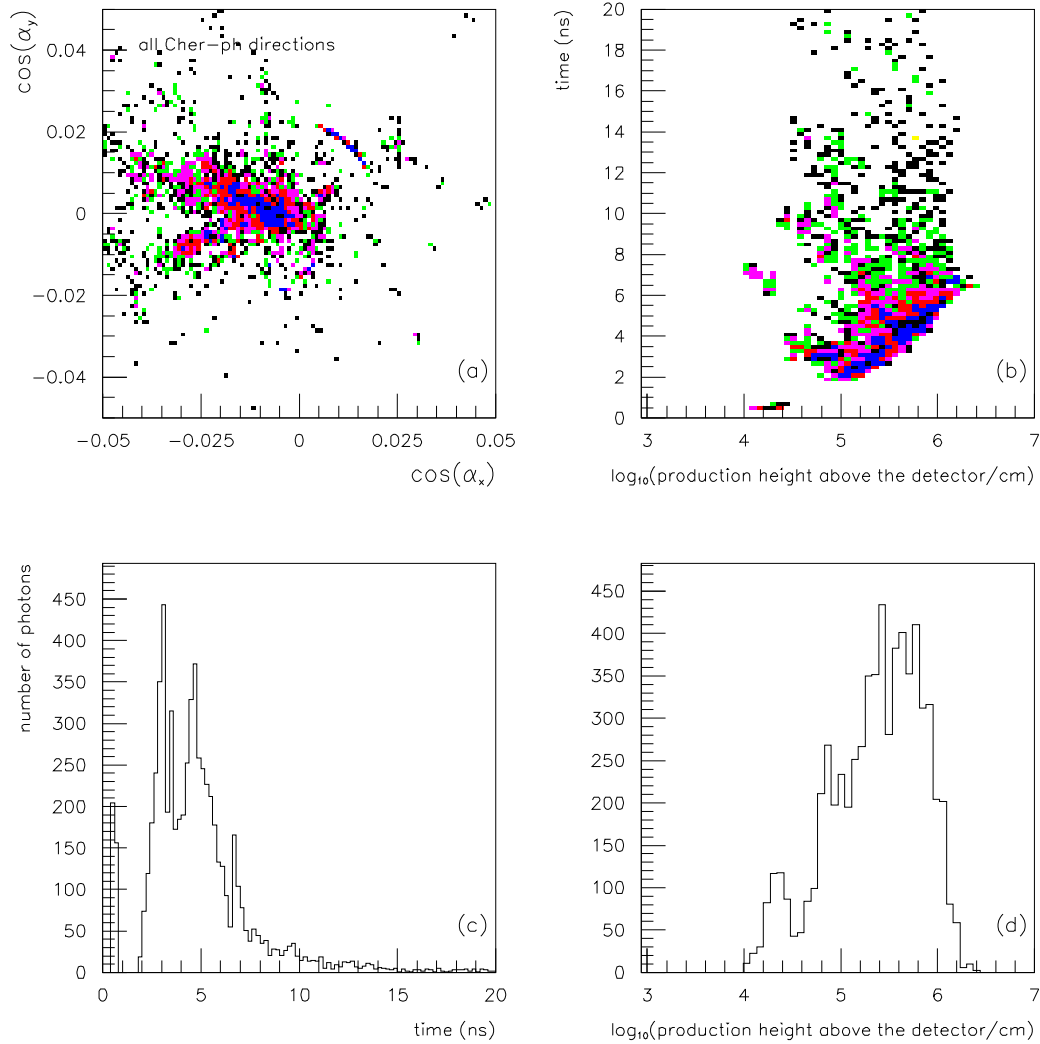,width=14.0cm,height=14.0cm}
 \odsrys\ 
 \caption
 {\label{fig:all_ph}
 Arrival time and production height distributions
 and correlations for Cherenkov photons produced
 (no absorption) in 10~TeV proton EAS (photons
 on 3~m~x~3~m area 50~m away from EAS core).
 See text for more comments and explanations.
 }
 \end{figure}
 \odstext\ 
 
 \noindent
 In the Figure~\ref{fig:all_ph}a 
 we present the angular distribution of Cherenkov
 photons 
 which fall on the area 3~m~x~3~m centred at (--50~m,~0~m)
 and are
 produced in the EAS described above.
 This is a 2--dimensional distribution for 
 cos($\alpha_{x}$) and cos($\alpha_{y}$) at the axes
 and grey scale corresponding to the decimal logarithm
 of the number of photons in the pixel (0.001~mrad~x~0.001~mrad).
 White colour indicates no photons in the pixel.
 We have $\theta_{\qqq{x}} = \pi /2 - \alpha_{\qqq{x}}$
 and $\theta_{\qqq{y}} = \pi /2 - \alpha_{\qqq{y}}$.
 For small zenith angle $\theta$, 
 cos($\alpha_{\qqq{x}}$) and cos($\alpha_{\qqq{y}}$)
 approximately correspond to 
 $\theta_{\qqq{x}}$ and $\theta_{\qqq{y}}$ in radians,
 the axes limits are from --50~mrad to +50~mrad in x and y
 directions.\\
 The result shown in the Figure~\ref{fig:all_ph}a
 corresponds to the theoretical registration of all photons  
 produced in EAS by
 the imaging Cherenkov device 
 at 50~m away from the EAS core.
 The main pattern has an assymetry which corresponds
 to the fact that most of photons are coming from the
 direction of the EAS axis (negative cos($\alpha_{\qqq{x}}$) values).
 The perpendicular spread of photon directions
 (in this case along cos($\alpha_{\qqq{y}}$) axis)
 is quite large, and this fact is often used to
 discriminate between gamma and hadron primary particle.\\
 Around the point (0.015,0.015) there is an arc of Cherenkov
 photons produced by the muon. The angular radius of the arc
 is about 20~mrad which indicates the relativistic particle,
 the centre of arc points to the direction of the muon
 mentioned above (-0.00196,0.00088).\\

 \noindent
 In the Figure~\ref{fig:all_ph}b
 we present the distribution of Cherenkov photons
 in 2--dimensional space of arrival time vs. height of
 photon emission (these are the same photons
 as in the Figure~\ref{fig:all_ph}a).
 Most of photons are produced 600~m above the detector level
 or higher and arrive with $\Delta t~>$~2~ns. 
 The small spot at place $\Delta t$~=~0.5~--~1~ns and
 production height 120~--200~m corresponds to the
 Cherenkov light emitted by the energetic muon.
 The isolated spot with $\Delta t$~=~6~--~7~ns and
 production height 120~--~200~m corresponds to another,
 not very fast muon.\\

 \noindent
 In the Figure~\ref{fig:all_ph}c
 we show the
 arrival time distribution
 (projection of the Figure~\ref{fig:all_ph}b
 on the time axis)
 The first peak in the histogram represents
 Cherenkov photons emitted by the energetic
 muon.\\
 In the Figure~\ref{fig:all_ph}d
 there is production height distribution
 (projection of the Figure~\ref{fig:all_ph}b
 on the height axis). 
 The bump around 4.3~$\approx$~log$_{10}$(200~m)
 is due to the accumulated signals from muons.
 From all charged 

 \begin{figure}[hb]
 \psfig{file=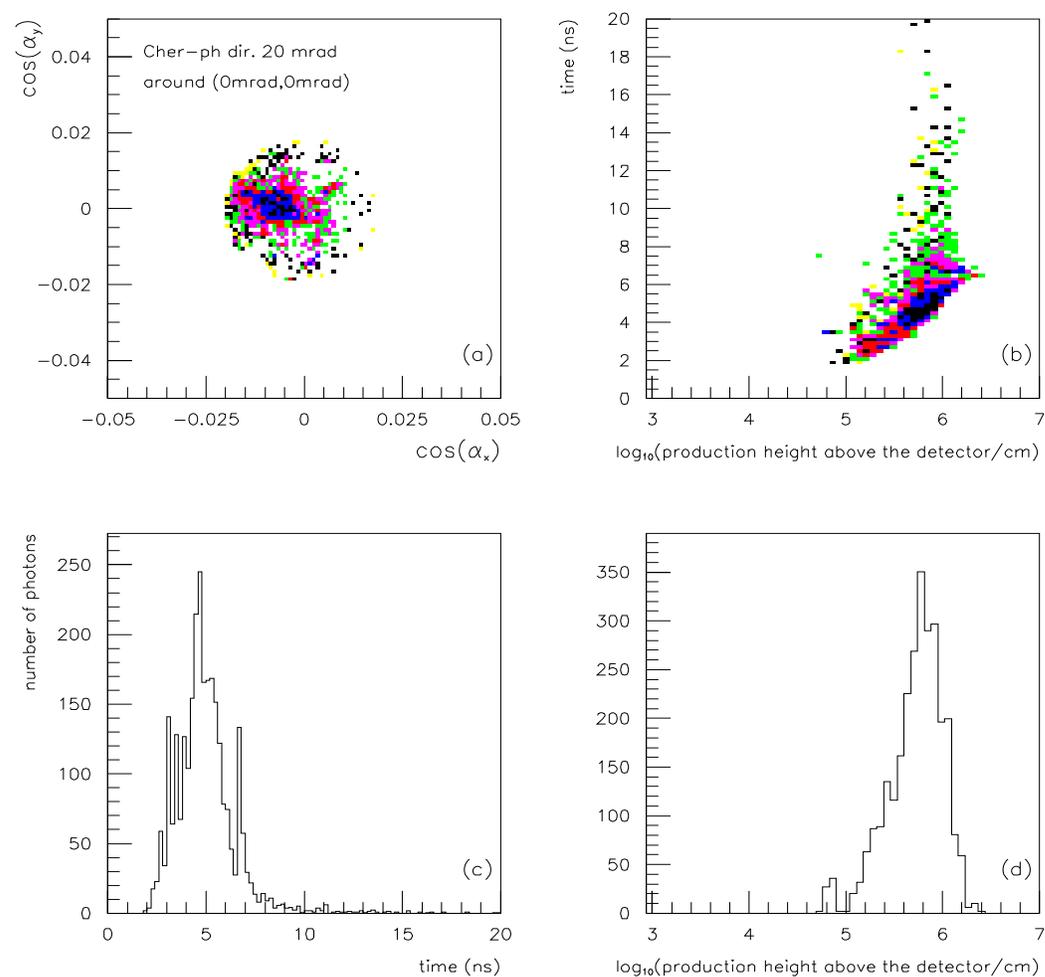,width=14.0cm,height=13.0cm}
 \odsrys\ 
 \caption
 {\label{fig:0:0}
 Similar to Figure~\ref{fig:all_ph} but
 only photons within 20~mrad from vertical direction.
 Arrival time and production height distributions
 and correlations for Cherenkov photons produced
 (no absorption) in 10~TeV proton EAS (photons
 on 3~m~x~3~m area 50~m away from EAS core).
 See text for more comments and explanations.
 }
 \end{figure}
 \odstext\ 
 
 \noindent
 \parbox{8.5cm}{
 particles of 10~TeV proton EAS
 only muons can get down below 2~km a.s.l. 
 (i.e. around 200~m above the detector placed
 at altitude 1650~m a.s.l.).\\
 Let us demonstrate a role of angular acceptance
 of each detector. This means that not all Cherenkov
 photons which are reflected from the mirros, fall
 onto the cathode of the phototube.
 In the case of the THEMISTOCLE heliostat
 the angular acceptance can be described as a 
 probability function $P(\gamma)$, where $\gamma$ is
 the angular difference between the direction of photon
 and optical axis of the mirror:\\
 \[
 P(\gamma) \, = \,
 \left\{
 \begin{array}{ll}
 1 & {\rm if~} \, \gamma < {\rm 10~mrad} \\ 
  & \\
 2 - \frac{\qqq{\gamma}}{\qqq{\rm 10~mrad}} & 
      {\rm if~} \, {\rm 10~mrad} < \gamma < {\rm 20~mrad} \\ 
  & \\
 0 & {\rm if~} \, \gamma > {\rm 20~mrad} 
 \end{array}
 \right.
 \]
 }
 \hfill
 \begin{minipage}{6cm}

 \setlength{\unitlength}{1mm}
 \begin{picture}(60,50)
 \put(10,10){\line(0,1){35}}
 \put(10,10){\line(1,0){45}}
 \put(10,40){\line(1,0){15}}
 \put(25,40){\line(1,-2){15}}
 \put(25,10){\line(0,1){2}}
 \put(40,10){\line(0,1){2}}
 \put(23,6){\makebox(0,0)[bl]{\it 10~mr}}
 \put(38,6){\makebox(0,0)[bl]{\it 20~mr}}
 \put(6,40){\makebox(0,0)[bl]{\it 1}}
 \put(6,10){\makebox(0,0)[bl]{\it 0}}
 \end{picture}
 \odsrys\ 
 \addtocounter{figure}{1}
 Figure~\thefigure : 
 Approximate THEMISTOCLE experiment detector
 acceptance angle distribution 
 ({\em P.~Espigat, 1995, private communication})\\
 \odstext\ 

 \end{minipage}
 \\
 \begin{figure}[hb]
 \psfig{file=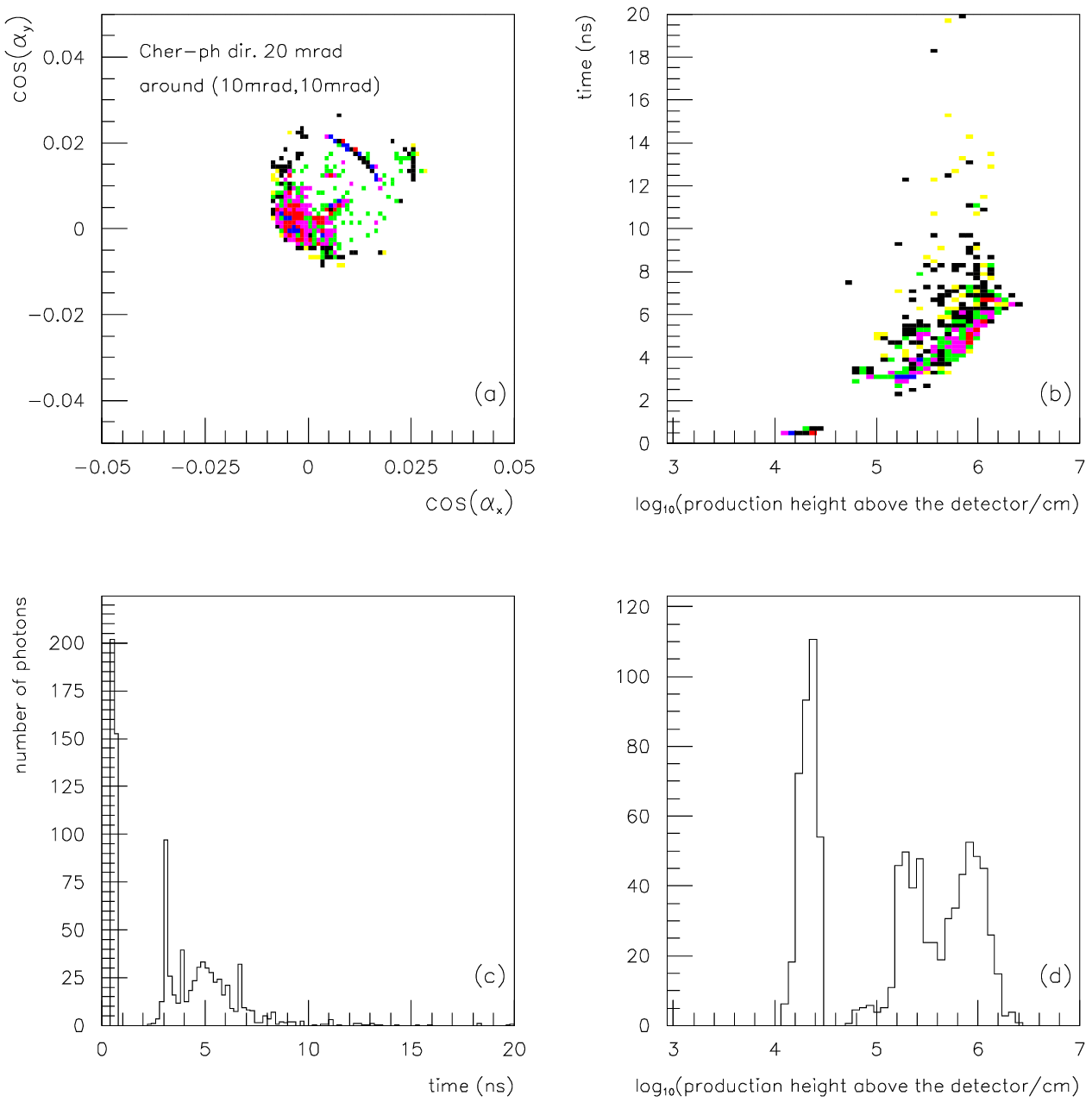,width=14.0cm,height=14.0cm}
 \odsrys\ 
 \caption
 {\label{fig:10:10}
 Similar to Figure~\ref{fig:all_ph} but
 only photons within 20~mrad around the direction 
 ($\theta_{\qqq{x}},\theta_{\qqq{y}}$) equal to
 (10~mrad,~10~mrad).
 Arrival time and production height distributions
 and correlations for Cherenkov photons produced
 (no absorption) in 10~TeV proton EAS (photons
 on 3~m~x~3~m area 50~m away from EAS core).
 This time the muon signal is clearly seen.
 See text for more comments and explanations.
 }
 \end{figure}
 \odstext\ 
 
 \noindent
 In the Figure~\ref{fig:0:0} 
 we present a \lq subset' of photons shown in 
 the Figure~\ref{fig:all_ph} 
 taking into account the THEMISTOCLE mirror acceptance
 for the mirror optical axis in the direction of EAS,
 i.e. directional cosines (0.,0.).
 Since the muon direction was nearly vertical,
 the muon pattern shown
 in the Figure~\ref{fig:all_ph}a
 disappeared 
 in the Figure~\ref{fig:0:0}a 
 (the Cherenkov cone angle is just 20~mrad).
 The muon signal is absent in the
 Figure~\ref{fig:0:0}b; we have 
 only Cherenkov light from the E--M component of EAS
 present.
 The fastest photons arrive with 
 $\Delta t~\approx$~2~ns and the 
 \lq advanced' peak disappeared from
 the Figure~\ref{fig:0:0}c.
 The small bump seen 
 in Figure~\ref{fig:0:0}d
 at 800~m above the observation
 level (log$_{10}$(height/cm)~$\approx$~4.8)
 is probably due to another muon
 which produced Cherenkov photons with
 $\Delta t~\approx$~4~ns 
 (see Figure~\ref{fig:0:0}b).\\

 \noindent
 In the Figure~\ref{fig:10:10} 
 we present another \lq subset' of photons shown in 
 the Figure~\ref{fig:all_ph}.
 This time the mirror \lq points' to the
 direction (10~mrad,~10~mrad) relative to the EAS
 direction. This direction has been selected to
 present clearly the signal of muon.
 The area of the detector is still 3~m~x~3~m.\\
 In the Figure~\ref{fig:10:10}a 
 the arc of Cherenkov photons generated by muon
 is clearly seen. The muon signal is also
 exposed on other Figures~\ref{fig:10:10}b, c and d.
 We would pay special attention to the
 Figure~\ref{fig:10:10}c which presents
 $\Delta t$ distribution, since it is 
 potentially possible to be measured.\\
 \begin{figure}
 \psfig{file=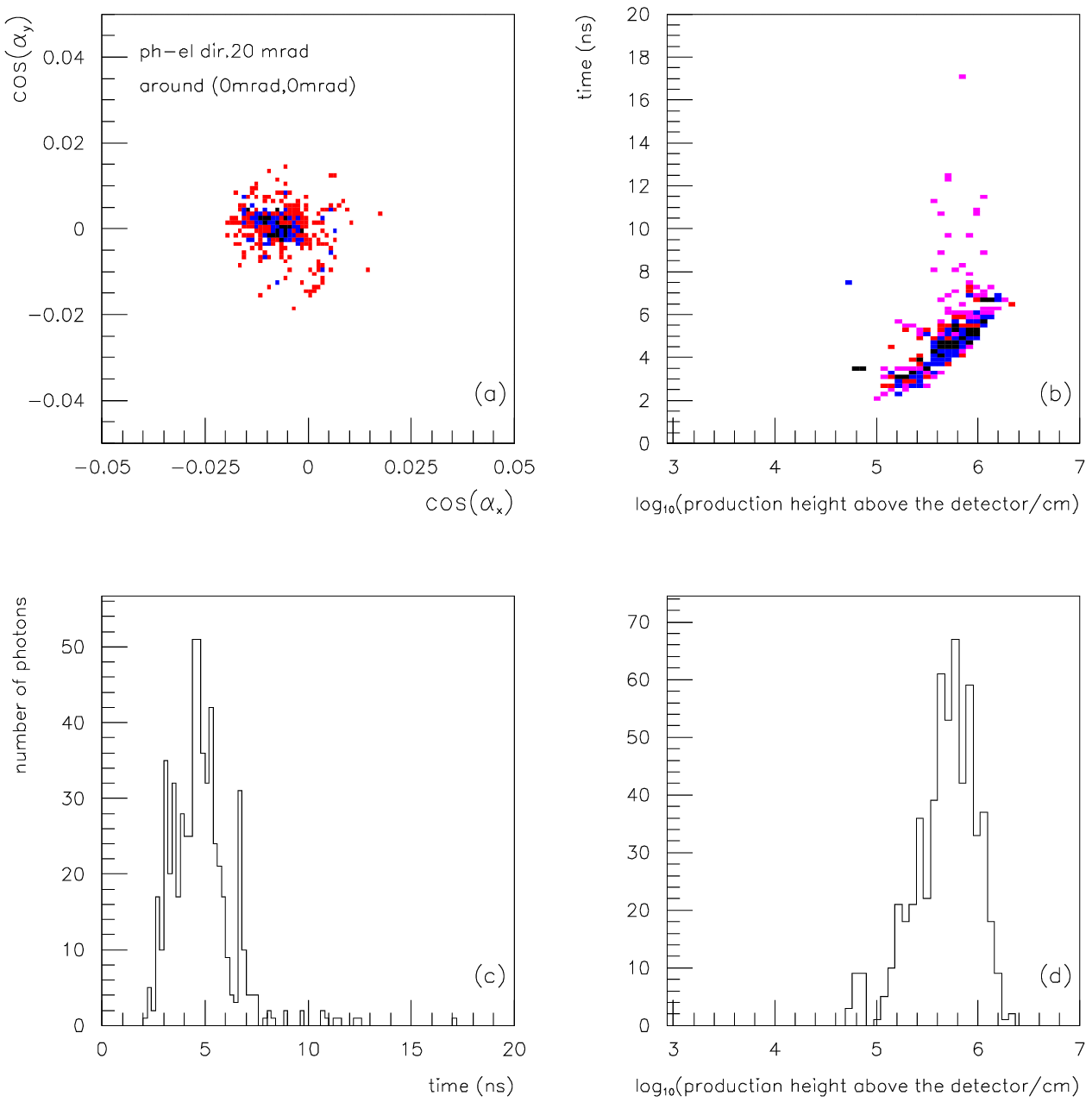,width=14.0cm,height=14.0cm}
 \odsrys\ 
 \caption
 {\label{fig:pe:0:0}
 Similar to Figure~\ref{fig:0:0} but
 for photo--electrons (and corresponding photons);
 only ph--el from photons within 20~mrad around 
 the vertical direction. 
 Arrival time of ph--el and production height distributions
 and correlations for related Cherenkov photons
 in 10~TeV proton EAS 
 (detector area of 3~m~x~3~m, 50~m away from EAS core).
 See text for more comments and explanations.
 }
 \end{figure}
 \odstext\ 
 \begin{figure}[tb]
 \psfig{file=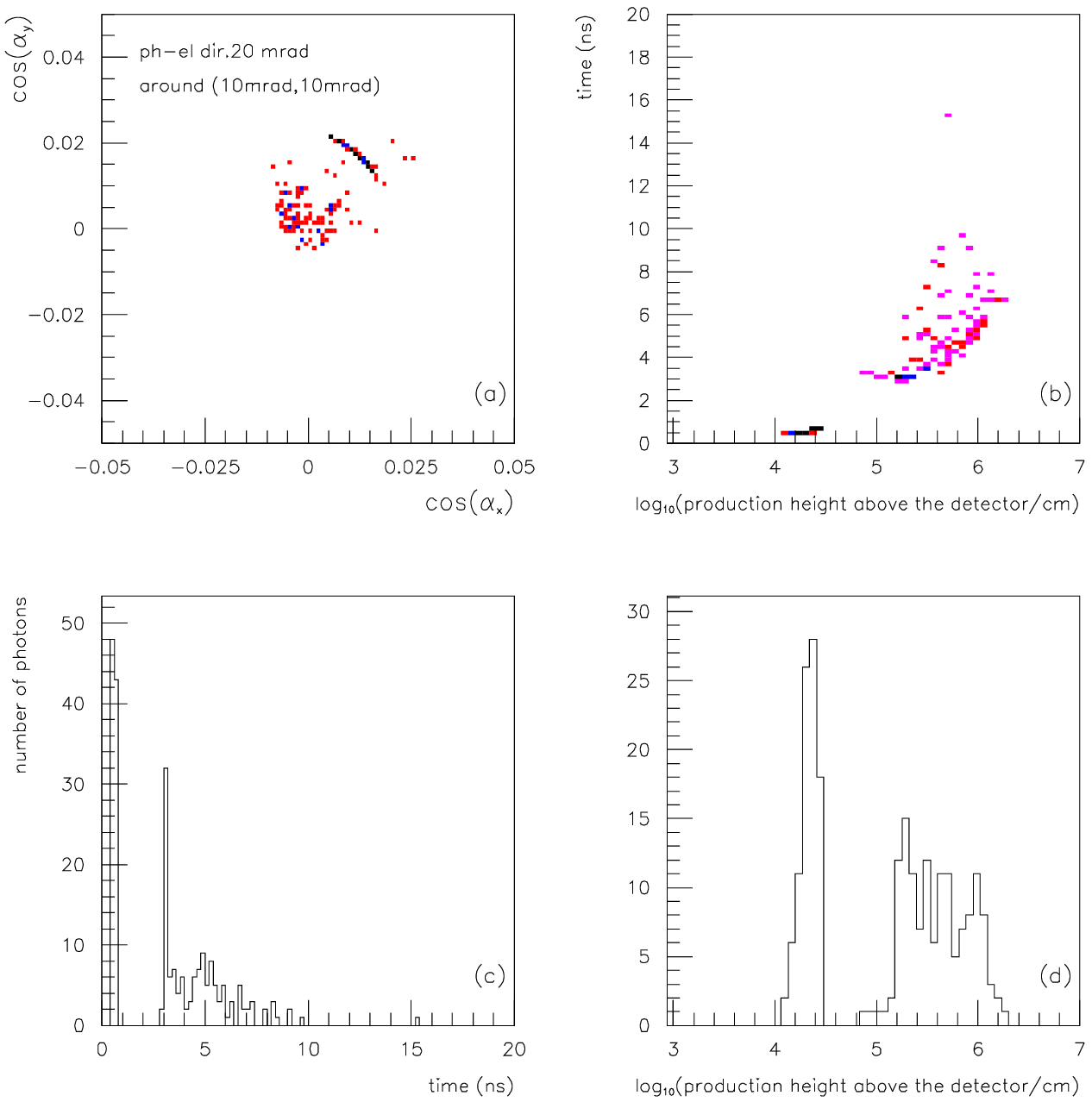,width=14.0cm,height=14.0cm}
 \odsrys\ 
 \caption
 {\label{fig:pe:10:10}
 Similar to Figure~\ref{fig:10:10} but
 for photo--electrons (and corresponding photons);
 only ph--el from photons within 20~mrad around 
 the direction (10~mrad,~10~mrad).
 Arrival time of ph--el and production height distributions
 and correlations for related Cherenkov photons
 in 10~TeV proton EAS 
 (detector area of 3~m~x~3~m, 50~m away from EAS core).
 This time the muon signal is clearly seen.
 See text for more comments and explanations.
 }
 \end{figure}
 \odstext\ 
 
 \noindent
 In the real case we need to consider 
 the photo--electrons from the photomultiplier
 cathode instead of Cherenkov photons produced.
 The processes of photon absorption, mirror
 reflection and photomultiplier quantum efficiency
 were included in the Monte--Carlo program
 \cite[{\em Attallah et al., 1995}]{Czer_I}, 
 which 
 processed the results of CORSIKA program 
 (Cherenkov photons in 300--450~nm).
 In the Figure~\ref{fig:pe:0:0} 
 we show time distributions of photo--electrons 
 and angular and production height distributions 
 of photons corresponding to photo--electrons
 for the 3~m~x~3~m detector at (--50~m,~0~m)
 pointing vertically
 with angular acceptance within 20~mrad
 (related to the Figure~\ref{fig:0:0}).
 The difference due to transition from
 the \lq emitted photon' case to 
 the \lq photo--electron' case can be
 seen by comparison of ordinate axes
 between Figures~\ref{fig:0:0} and \ref{fig:pe:0:0}
 for c) and d) histograms.\\

 \noindent
 In the Figure~\ref{fig:pe:10:10} 
 we present the photo--electron distributions
 for detector pointing to (10~mrad,~10~mrad)
 direction to the EAS direction and including
 the muon signal
 (related to the Figure~\ref{fig:10:10}).
 For the discussed problem of Cherenkov light
 from fast muons the 2~ns gap seen 
 in the Figure~\ref{fig:pe:10:10}c is very important.\\

 \noindent
 The Figures~\ref{fig:pe:0:0}c and~\ref{fig:pe:10:10}c
 ($\Delta t$ distributions for photo--electrons)
 differ quite significantly.
 However, photomultiplier anode signal 
 does not reproduce these patterns accurately.
 Here we present results of signal convolution
 for two photomultipliers: PHILIPS~XP2020 and
 Hamamatsu~H2083. 
 (The XP2020 data after
 \cite[{\em Philips Handbook, 1990}]{PhilipsHand}
 and
 Hamamatsu~H2083 after
 \cite[{\em Riera and Espigat, 1994}]{Riera}).
 We use following parametrization
 of the anode impulse corresponding to one photo--electron
 \cite[{\em PHILIPS Composants, 1990}]{PH:tubes}:
 \begin{equation}
 R_{\delta}(t) = G \, \cdot \,
 \frac{\qqq{\sqrt{m\, + \, 1}}}{\qqq{m! \, \sigma_{\rm R}}}
 \, \cdot \,
 \left(
 \frac{\qqq{\sqrt{m\, + \, 1}}}{\qqq{m! \, \sigma_{\rm R}}} \, \cdot t \,
 \right)^{\qqq{m}}
 \, \cdot \,
 exp\left( -
 \frac{\qqq{\sqrt{m\, + \, 1}}}{\qqq{m! \, \sigma_{\rm R}}} \, \cdot t \,
 \right)
 \label{eq:PMsignal}
 \end{equation}
 where $R_{\delta}(t)$ is the time dependent (negative) anode signal
 corresponding to $\delta$ impulse -- one photo--electron,
 $G$ is the actual gain (current amplification),
 $t$ is time (see below for more details),
 $m$, $\sigma_{\rm R}$ are parameters related to the pulse shape.
 The $R_{\delta}(t)$ has following properties:
 \[
 {\displaystyle
 \int_{\qqq{0}}^{\qqq{\infty}} \, R_{\delta}(t) \, dt \,
 = \, G
 }
 \] \[
 t_{\qqq{max}} \, = \, 
 \frac{\qqq{m \, \sigma_{\rm R}}}{\qqq{\sqrt{m\, + \, 1}}}
 \] \[
 R_{\qqq{max}} \, = \, R_{\delta}(t_{\qqq{max}}) \, = \, G \, \cdot \,
 \frac{\qqq{\sqrt{m\, + \, 1}}}{\qqq{m! \, \sigma_{\rm R}}}
 \, \cdot \,
 m^{\qqq{m}} \, \cdot \, e^{\qqq{-m}}
 \]
 The time between the photo--electron emission from the
 cathode and time of the maximum of signal $t_{\qqq{max}}$
 is called the transit time and its average value depends on
 photomultiplier, high voltage and divider.
 In this calculation we set the average transit time
 to be 30~ns for both photomultipliers (although it is different
 in the real case, e.g. about 39~ns for XP2020).
 Jitter is the variability of that time. Jitter has the
 Gaussian distribution with $\sigma_{\qqq{t\_ jitter}}$~=~0.25~ns.
 We assume another fluctuation of transit time
 due to the different photo--electron emission place
 in the cathode to be of Gaussian shape with 
 $\sigma_{\qqq{t\_ cat}}$~=~0.25~ns.\\
 \begin{figure}[tb]
 \psfig{file=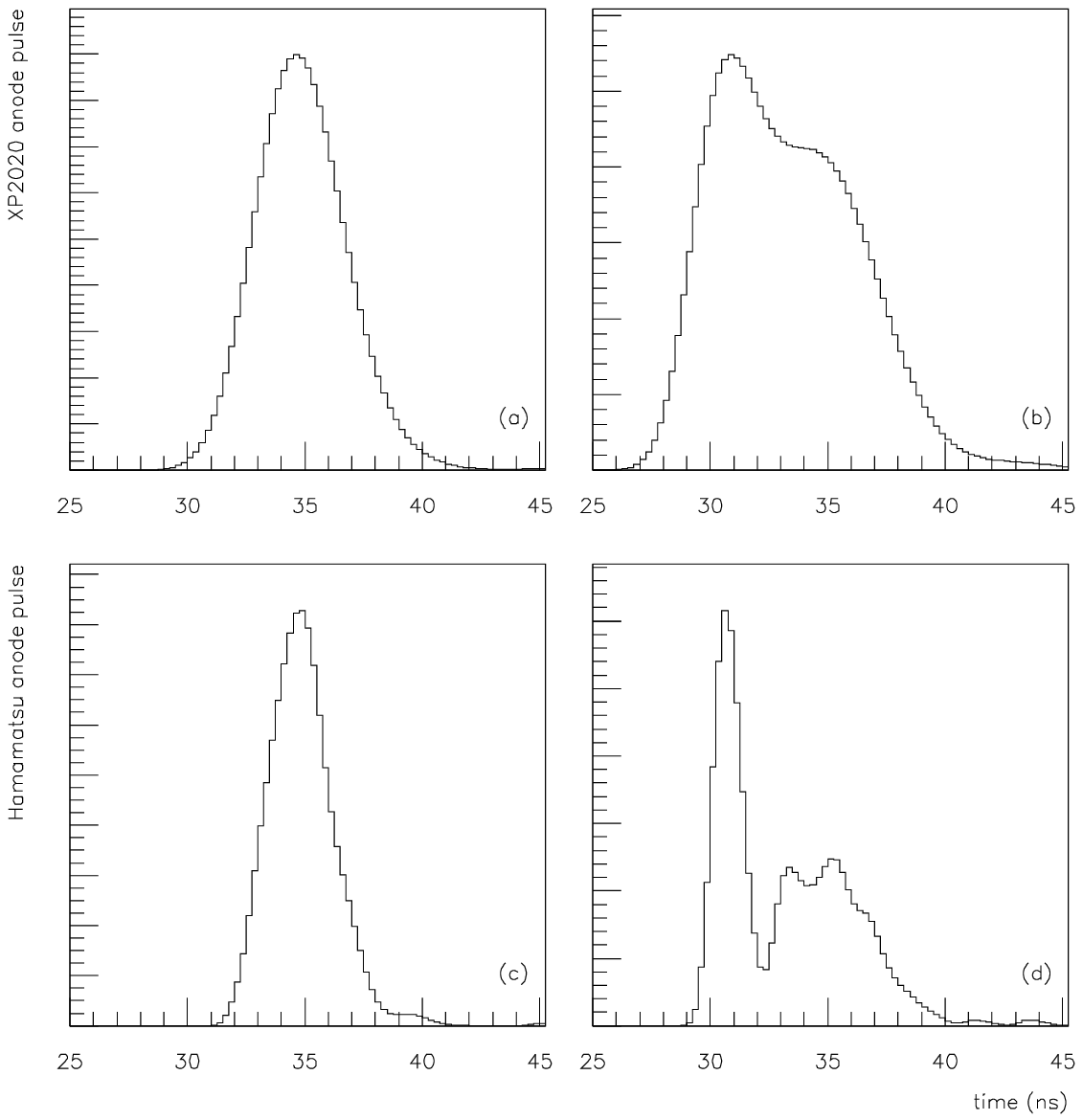,width=12.0cm,height=12.0cm}
 \odsrys\ 
 \caption
 {\label{fig:anode}
 The anode signal (arbitrary units)
 for photomultipliers:
 PHILIPS~XP2020 (~a and b at the top)
 and
 Hamamatsu~H2083 (~c and d at the bottom)
 for 3~m~x~3~m detector at 50~m away from 
 the EAS core. Left figures (~a and c)
 correspond to photo--electron distribution
 shown in the Figure~\ref{fig:pe:0:0} (no muon signal),
 right figures (~b and d)
 to the Figure~\ref{fig:pe:10:10} with
 Cherenkov light from the muon $\sim$~2~ns
 before the E--M Cherenkov light signal.
 }
 \end{figure}
 \odstext\ 
 \begin{tabular}{llllllrr}
  & voltage & Gain & $\sigma_{Gain}$ & 
     FWHM & rise time & 
     $m$ & $\sigma_{\rm R}$ \\
  & & & & & & & \\
  XP2020 & 1800~V & 2 \, $\cdot$ \, 10$^{\qqq{4}}$ & 0.35 $\cdot Gain$ &
     3.4~nsec & 2.1~nsec &
     50 & 1.44 \\
  & & & & & & & \\
  Hamamatsu~H2083 & 3000~V & 10$^{\qqq{6}}$ & 0.4  $\cdot Gain$ &
     1.2~nsec & 0.6~nsec &
     7 & 0.55 \\
  & & & & & & & \\
 \end{tabular} \\
 
 \noindent
 where rise time is the time between 10\% and 90\% of 
 maximum value,
 $m$ and $\sigma_{\rm R}$ are parameters of
 equation~\ref{eq:PMsignal}.\\

 \noindent
 In the Figure~\ref{fig:anode} 
 we present the simulated shapes of anode
 signals (reversed to positive)
 for photomultipliers  PHILIPS~XP2020 and
 Hamamatsu~H2083
 for input photo--electron distributions
 presented in  
 the Figures~\ref{fig:pe:0:0}c and~\ref{fig:pe:10:10}c. 
 Units are arbitrary, since, effectively, this is further
 electronics dependent.
 One can notice that at low discrimination level
 the signals with muon and E--M Cherenkov light 
 last longer than signals without muon.
 In both cases (i.e. XP2020 and Hamamatsu~H2083)
 the signal for the case with the fast muon
 in the field of view starts 2~ns earlier than
 the signal form E--M Cherenkov light.\\

 \noindent
 This fact (time of start of the signal in one detector)
 does not discriminate between the no--muon and muon cases.
 To see the effect of fast muon
 (examining the start time)
 it is necessary to examine the start time in 
 a number of detectors for the same event.
 In most cases (also experimental) it is possible
 to find the position of the EAS core by
 imposing the requirement of the cone alignement
 of starting time for detectors placed at various
 distances (see Figure~\ref{fig:delta_t}).
 (Such a method is used in the THEMISTOCLE 
 experiment, 
 \cite[{\em Baillon et al., 1993}]{THEMISTOCLE}.
 Since most of detectors do not have 
 Cherenkov light from fast muon in the field of view
 the alignement is usually very good.
 Then one detector with starting time several
 nsec in advance to the alignement obtained
 from other detectors might indicate the presence
 of the fast muon in the EAS, and
 such an event is almost surely of 
 primary CR hadronic origin.
 Viewing the THEMISTOCLE data such events
 have been found.\\
 However there is a problem with the night sky noise background,
 which might mimic the signal due to fast
 muon. It is not very likely to get 3~or more photo-electrons
 due to the night sky background within the
 XP2020 pulse width in the THEMISTOCLE experiment.
 The night sky background is
 about 1300~ph/(nsec~m$^{2}$~sr) (330--550~nm)
 \cite[{\em Dumora et al., 1996}]{Celeste},
 which produce on average $\sim$~0.04~ph--el/nsec in one THEMISTOCLE
 detector, 
 or 2~ph--el within 3~nsec every $\sim$~200~nsec.
 The timing electronics might (might not) indicate
 2--3~ph--el as a beginning of the event, but
 this depends on the amplitude of forecoming main signal.\\

 \noindent
 One of the possible solutions of the identification
 problem is to use fast Flash~ADC. Such a device
 would allow to sample the anode amplitude 
 every nsec. Using 
 Hamamatsu~H2083 
 and Flash~ADC one would expect to have
 information similar to that shown in the
 Figures~\ref{fig:anode}c and d.
 It would be possible to fit a cone timing shape
 to all detectors and to see the structure of
 the signal from the detector starting 
 several nanoseconds earlier relatively to its
 distance to the EAS core.\\

 \noindent
 Similar registrations have been performed for
 THEMISTOCLE experiment with one extra
 detector with 
 Hamamatsu~H2083 and Flash~ADC 
 ({\em P.~Espigat, private communication, 1996}).
 In the CELESTE experiment all fast photomultipliers
 would be equipped with the Flash~ADC.
 \begin{figure}[tb]
 \psfig{file=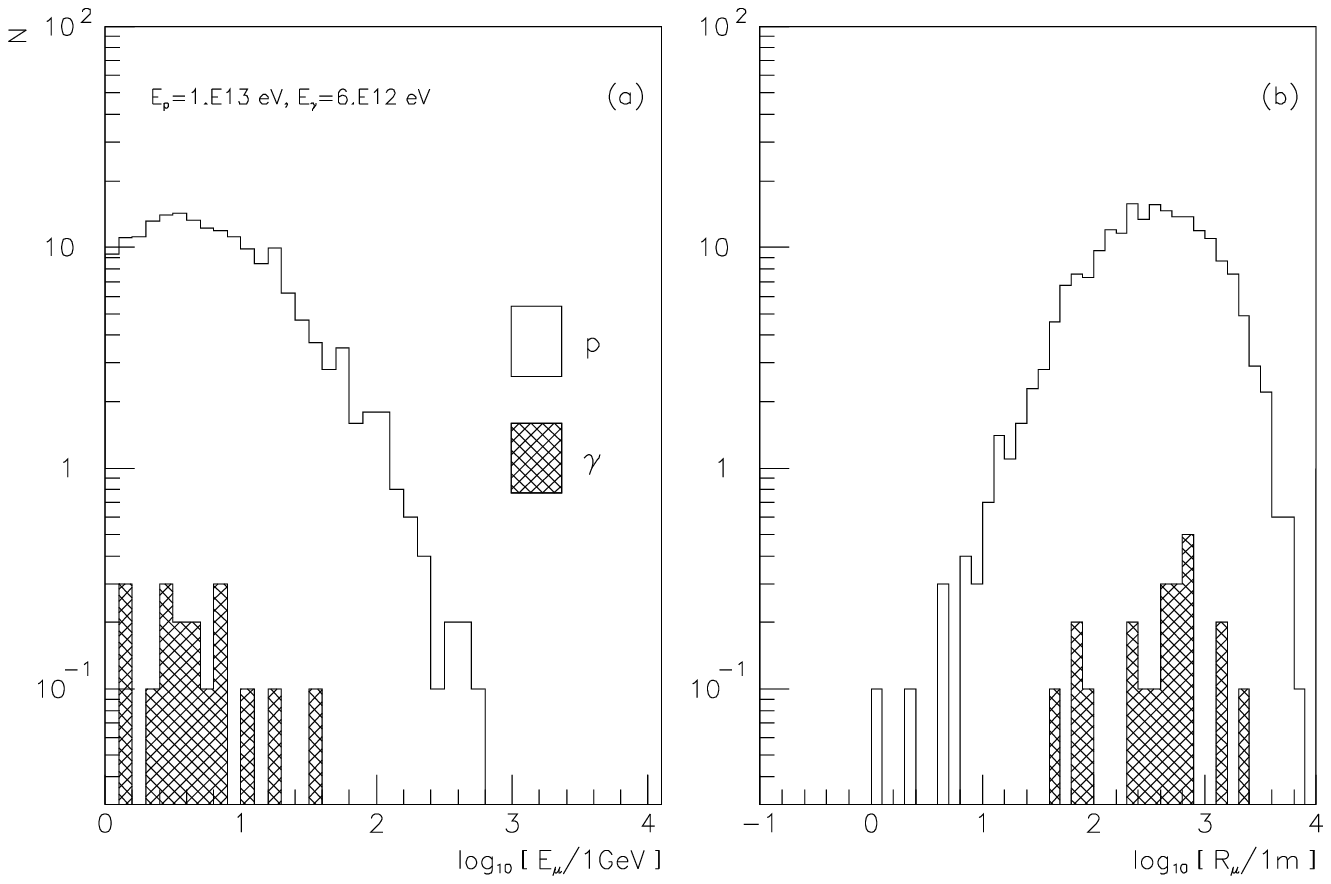,width=14.0cm,height=9.0cm
 }
 \odsrys\ 
 \caption
 {\label{fig:muons}
 The average muon energy and lateral distributions
 for 10 vertical EAS produced by 10~TeV CR protons
 and 6~TeV~CR~$\gamma$'s as seen    
 at altitude 1750~m~a.s.l. (860~g/cm$^{2}$).
 }
 \end{figure}
 \odstext\ 
 \vspace{-10pt}
 \section{
 Muons in TeV EAS.
 }
 Muons in EAS are decay products of pion and kaons.
 The probability of decay depends on the lifetime
 of hadron, its energy/mass ratio (factor~$\gamma$)
 and the density of the atmosphere (hadrons might
 interact or decay).
 When the primary CR particle is a proton or nucleus
 then hadron production is a natural consequence
 of its inelastic interaction while passing through
 the atmosphere.
 When the primary CR particle is of E--M type
 (i.e. e$^{+}$, e$^{-}$ or $\gamma$)
 then hadrons are produced due to 
 photoproduction process.\\

 \noindent
 Here we present the average distribution
 of muons produced in 10~TeV proton EAS and 
 6~TeV~$\gamma$ EAS both from vertical direction. 
 The energies of primary particles were selected
 to produce similar Cherenkov photon densities
 at 50~m from the EAS core.
 In the Figure~\ref{fig:muons}a we present
 average energy distribution 
 (i.e. $dN_{\mu}/d(log_{10}(E_{\mu}/GeV))$).
 For primary CR proton the number of muons
 with energy above 1~GeV is $\approx$~120,
 above 5~GeV is $\approx$~40 (they do produce
 Cherenkov light), and 
 above 10~GeV is $\approx$~20 per 1~EAS.
 For primary CR $\gamma$ the number of muons
 is approximately 100~times lower.\\
 The average lateral distribution
 of muons 
 (Figure~\ref{fig:muons}b~)
 shows that the muon density is nearly
 constant till 100~m away from the EAS core 
 (the $dN_{\mu}/d(log_{10}(R/1~m))$ has a power law
 dependence with exponent nearly equal to 2).
 Muons which origin near the EAS core on $\sim$10~km
 and come about 100~m from the EAS core on the ground
 are inclined by $\sim$10~mrad. Since the Cherenkov cone
 is $\sim$20~mrad near the ground, part of the Cherenkov light
 produced by them would be registered when EAS come from the
 observation direction.

 \vspace{-10pt}
 \section{
 Conclusions.
 }
 In this paper we have discussed the possibility
 to distinguish between electromagnetic EAS
 and hadronic EAS in \lq wavefront sampling'
 Cherenkov light measurements of TeV cosmic rays.
 The main target of those experiments is to
 identify and measure TeV $\gamma$ flux
 from selected sources. 
 The positive measurements would increase our 
 knowledge about nature of high energy particle sources.\\
 Positive results obtained so far
 identify the excess of source DC flux
 by comparison with the \lq off source' measurements.
 For example, in the THEMISTOCLE experiment
 \cite[{\em Baillon et al., 1993}]{THEMISTOCLE}
 the observation of Crab nebula
 gained about 1440 events localized within 5~mrad
 from the Crab pulsar direction. From 
 \lq off source' observation the expected background
 within 5~mrad was estimated.
 The excess $\approx$~260 events are the Crab nebula DC signal.
 The rest of $\approx$~1180 events are mostly due to hadronic
 EAS background.\\
 Using any other method of identifying the hadronic nature
 of observed event it would be possible to
 increase signal to noise ratio and 
 add the strength to the result observed so far.
 We presented here some details of differences 
 between E--M EAS and hadronic EAS due to the presence
 of fast muons and Cherenkov light produced by them.\\
 It is very difficult to give a quantitative calculated
 prediction about the efficiency of this method.
 There are many problems and only some of them were addressed
 here. We hope that examining carefully already existing
 experimental data one would be able to see a muon
 signal and then estimate the efficiency of the method
 in the specific experimental conditions.

 \vspace{20pt}
 \noindent
 \underline{
 \bf
 Acknowledgements.}
 We would like to thank Dr. Pierre Espigat and Dr. Claude Ghesqui\`{e}re
 for very valuable discussions.

 \end{document}